\documentclass[flushrt]{aastex}
\usepackage{emulateapj5}

\newcommand{\etal}{{\it et~al.\/}\ }
\newcommand{\minusone}{$^{-1}$}

\newcommand{\II}{\protect\small II \normalsize $\!\!$}
\newcommand{\HII}{\mbox{\rm H\II}\ }

\shorttitle{A Search for Methanol Masers in OH Megamaser Galaxies}
\shortauthors{Darling, Goldsmith, Li, \& Giovanelli}

\begin{document}
\title{A Search for 6.7 GHz Methanol Masers in OH Megamaser Galaxies at 
$0.11<z<0.27$}
\author{Jeremy Darling\altaffilmark{1}, 
	Paul Goldsmith\altaffilmark{2}, 
	Di Li\altaffilmark{3}, 
	\& Riccardo Giovanelli\altaffilmark{2}}
\altaffiltext{1}{Carnegie Observatories, 813 Santa Barbara Street, Pasadena, 
		 CA  91101}
\altaffiltext{2}{Cornell University and National Astronomy and Ionospheric 
		 Center, Ithaca, NY 14853}
\altaffiltext{3}{Smithsonian Astrophysical Observatory, 60 Garden Street, 
		 MS 66, Cambridge, MA 02138}

\begin{abstract}
We report the results of a search for 6.7 GHz methanol (CH$_3$OH) maser 
emission in OH megamaser galaxies at $0.11<z<0.27$.  No detections were 
made in the 25 objects observed down to rms noise levels of 
roughly 0.6 mJy in 150 kHz channels.  
The OH megamaser sample includes OH emission of all observed types:
narrow and broad, physically compact and extended, variable and quiescent,
and from simple single lines to 
multi-component complexes to lines with high velocity wings.
Our null result indicates,
for the widest possible range of OH megamaser environments, 
that methanol masing does not scale with OH from Galactic masers to 
megamasers.  These observations, however, are not sensitive enough to rule
out methanol masing that scales with star formation from Galactic 
compact \HII\ regions to starbursts associated with major mergers.  
This and previous work suggest that OH megamasers do not represent large 
ensembles of individual masers associated with star forming regions.
Maser models combined with observational constraints on the physical
settings of OH megamasers indicate that 6.7 GHz methanol megamasers will
not be detected by this survey if $T_{dust}<100$ K or if
$n_{CH_3OH} \leq n_{OH}$.
\end{abstract}

\keywords{masers --- galaxies:  ISM 
--- ISM: molecules --- galaxies: starburst --- radio lines: galaxies}

\section{Introduction}
A survey for OH megamasers (OHMs) in luminous IR galaxies has recently
been completed at the Arecibo Observatory\footnote{The Arecibo Observatory 
is part of the National 
Astronomy and Ionosphere Center (NAIC), which is operated by Cornell
University under a cooperative agreement with the National Science 
Foundation.} \citep{dar02a}.
The survey detected 50 previously unknown OHMs at $z=0.1$--0.25, more
than doubling the sample of OHMs overall, and increasing the sample 
at $z>0.1$ sevenfold.  OHMs are associated with starburst nuclei
in major galaxy mergers.  
The primary factors responsible for producing
OHMs appear to be the large-scale density enhancements 
of molecular gas produced exclusively in galaxy-scale tidal interactions 
and the strong far-IR radiation field which inverts the OH ground state, 
providing the maser pump.  If this is the
case, then the large-scale density enhancements of excited molecular gas 
combined with the intense IR radiation field can 
be expected to produce masing in other lines and other molecules.  We focus
here on the 6.7 GHz ($5_1\rightarrow6_0$ A$^+$) transition of methanol 
(CH$_3$OH).
This strong masing line is associated with compact \HII\  regions and 
OH masing \citep{men91}.
Galactic OH and CH$_3$OH masing regions identify \HII\  regions,
are likely to be commingled, and have similar lifetimes \citep{cas95a,cas95b}.
Typically, 6.7 GHz methanol emission is an order of magnitude stronger than
1665 MHz OH and has a similar width but differs in the detailed line structure
\citep{cas95a}.  Hence, if the Galactic relationship between OH and CH$_3$OH 
emission in compact \HII\  regions scales to megamaser luminosities, 
then one would expect the 6.7 GHz line be easily detectable at cosmological
distances.  If, however, OH megamasers represent a fundamentally different 
physical setting from Galactic OH masers
then the scaling would not necessarily apply.  

Detections of extragalactic methanol masers are limited to the Large 
Magellanic Cloud \citep{sin92,ell94b,bea96}, 
with isotropic integrated line luminosities similar to observed
Galactic methanol masers (roughly 10$^{-5}$--10$^{-6}$ $L_\odot$).  
There are two surveys
for methanol megamasers described in the literature: both select OH
and H$_2$O megamaser hosts to search for 6.7 GHz methanol masers, and both
obtain only upper limits on maser emission \citep{ell94a,phi98}.
\citet{phi98} also surveyed a number of IR-luminous starburst
galaxies and made no detections.  
Does this paucity of detections indicate different mechanisms
for OHMs from the OH masers associated with 
Galactic star formation, or
are are methanol masers only associated with a subset of OHMs?  

This paper presents the results of a sensitive survey for 6.7 GHz methanol 
masers in OH megamaser hosts conducted at the Arecibo 
Observatory.
No detections were made, but the survey sets strong limits on the 
methanol emission for a wide variety of OH megamasers and explores a new
region of the redshift-luminosity space available to these extremely rare
objects.  The sample selection, observation methods, and data reduction
techniques are presented in \S \ref{data}.  Section \ref{results} presents
the limits on 6.7 GHz methanol maser emission in the 25 OHMs observed, 
and \S \ref{discussion} offers a brief interpretation of the implications
of this survey for methanol megamasers and the nature of OHMs.

\section{Data}\label{data}
\subsection{Sample Selection}  
A sensitive survey for 6.7 GHz methanol masers associated with OH megamasers 
would optimally include all known OHMs within the Arecibo declination 
range ($0^\circ<\delta<37^\circ$) up to the highest redshift OHM known at
$z=0.2655$ \citep{baa92}.  Unfortunately, the survey sample must be
restricted to $z\geq0.11$ for the redshifted 6.7 GHz methanol line to fall 
within the frequency range of the C-band receiver (3.95--6.05 GHz).  
No 6--8 GHz receiver is currently available at Arecibo.  

We select 28 OHMs at $z\geq0.11$ spanning the 
range of properties found in this diverse population.  Line widths 
range from unresolved to extremely broad (25--850 km s\minusone), 
peak flux densities span 2--40 mJy, and 1667:1665 MHz OH hyperfine 
ratios range from 2.5 to $>15$ (the LTE value is 1.8; see Table 
\ref{targets} and Darling \& Giovanelli 2002b).  The sample 
includes Seyferts type 1 and 2, LINERs, and pure starburst nuclei
and spans the range of (ultra)luminous IR galaxies ([U]LIRGs) 
on the radio-far-IR
relation, from radio loud to IR overluminous \citep{dar02a,dar03}.
The sample includes compact and extended OHMs as inferred from 
variability or quiescence of spectral features (Darling, in preparation).
The sample also includes OH lines with 
smooth profiles, indicating a distributed masing gas, lines with 
high velocity wings characteristic of molecular flows, and 
multiply-peaked profiles, indicating a wide range of masing conditions and
complicated kinematics.  
To maximize the chances of detection and to provide the strongest
possible constraint on the existence of methanol megamasers, 
the methanol maser survey sample includes every variety
of OHM and OHM host available.  This sample is complementary to the
\citet{phi98} survey which includes OH and H$_2$O megamasers and starburst
galaxies.

\subsection{Observations}
Observations were conducted (remotely) at the Arecibo Observatory with the
newly commissioned C-band (3.95--6.05 GHz) receiver on February 26--28, 
2002.  Each program source 
was observed in multiple 5-minute on-source, 5-minute off-source sessions
with a noise diode calibration at the end of each.  The radio frequency 
interference (RFI) environment at Arecibo generally requires fast sampling 
for RFI excision, but the effects of RFI at 4--6 GHz are relatively mild
and sampling every 6 s was adequate.
Spectra were produced using an autocorrelation spectrometer with 
9-level sampling in two linear polarizations.
We observed a 25 MHz band centered on the redshifted 6668.518 MHz 
$5_1\rightarrow6_0$ A$^+$ transition of CH$_3$OH flanked by 25 MHz bands 
offset by $\pm20$ MHz at the rest frequency and doppler tracked separately 
to detect any high velocity masing lines which might be present in the 
program sources.  Each 25 MHz bandpass had 1024 channels for
a spectral resolution of 1.3 km s\minusone\ and a total velocity coverage of 
3500 km s\minusone\ in the rest frame of a galaxy at $z=0.2$.  

The primary reflector was readjusted to $\sim2$ mm rms prior to 
these C-band observations, which improved the telescope forward gain
and the quality of the main beam.  
The beam size at 5 GHz is 54\arcsec$\times$63\arcsec\ (azimuth$\times$zenith
angle) and the pointing has a total rms error of about 5\arcsec.
The gain is about 8.5 K Jy\minusone\ at low zenith angles at 5 GHz, but is a 
strong function of zenith angle, falling to about 6.4 K Jy\minusone\ at
a zenith angle of 15$^\circ$ (the maximum zenith angle at Arecibo is 
19.7$^\circ$).  The gain is also a strong function of frequency, falling 
by roughly 20$\%$ from 5 GHz to 6 GHz.  The system temperature is nearly
flat at 34 K for zenith angles up to 15$^\circ$ at 5 GHz, but increases 
by $\sim15\%$ from 5 GHz to 6 GHz\footnote{See www.naic.edu for current 
receiver and system characterization information.}.
The methanol line for the program sources falls in the spectral range 
5270--6005 MHz, with the majority of the sources in the range 5470--5955 MHz.  


\subsection{Data Reduction}
All data reduction was performed in AIPS++\footnote{The AIPS++ (Astronomical 
Information Processing System) is a product of the AIPS++ Consortium.  AIPS++
is freely available for use under the Gnu Public License.  Further information
may be obtained from http://aips2.nrao.edu.} with the
NAIC single dish packages.  No RFI excision was done.  In the three objects
where RFI was significant, IRAS 06487+2208, IRAS 12162+1047, 
and IRAS 14586+1432, it was overwhelming and persistent and could not be
removed.  This RFI filled the range 5808--5840 MHz (42531--44417 km 
s\minusone).  On-off pairs and polarizations
were averaged to obtain the final spectra.   These spectra were Hanning smoothed,
and linear baselines were subtracted.  The bandpasses are generally excellent:
they show no more than a linear slope and are consistent across the 65 MHz 
of the three combined bandpasses.  

All spectra including the flanking fields were inspected after Hanning smoothing,
a 6-channel boxcar smoothing to 146 kHz, and gaussian smoothing matched to the
OH line width.  The boxcar smoothing nearly matches the 156 kHz boxcar-smoothed 
rms noise quoted by \citet{phi98}.  This is the noise level listed in column (9) of
Table \ref{targets}.  The observations are sensitive to methanol lines from 
a few to $\sim2000$ km s\minusone\ wide and would also identify 
any high velocity lines up to 1750 km s\minusone\ from the systemic redshift.

\section{Results}\label{results}
No credible emission or absorption lines were identified in the 25 RFI-free
program sources.
Table \ref{targets} lists the OHMs observed in the redshifted 6668.518 MHz methanol
line, including coordinates, optical heliocentric redshifts, OH line peak flux 
density, full width at half maximum, and isotropic line luminosity.  We assume 
a cosmology with $H_\circ = 75$ km s\minusone, $\Omega_M = 0.3$, and 
$\Omega_\Lambda = 0.7$ to compute luminosities.  To provide the most
conservative possible constraint on methanol emission relative to OH emission,
we compute the upper limit on the isotropic methanol line luminosity from
a 1.5$\sigma$ boxcar having width matching the OH line width $\Delta$v$_{OH}$,
where $\sigma$ is the rms noise of the 6-channel boxcar smoothed
spectrum.  We compute the same 6.7 GHz line luminosity limit for the 
\citet{phi98} OHM sample for confirmed OHMs with published widths (17 
objects), enabling direct comparison of the two samples.  This OH 
width-dependent methanol luminosity limit provides a direct measure of the
limit on the OH-to-methanol line luminosity under the assumption that the
two emission lines scale and are spatially coincident (this is
true in Galactic \HII regions; Caswell, Vaile, \& Forster 1995).  
The scaling of 
methanol masers with OH from Galactic \HII\ regions up to OH megamaser
scales is a straw man
hypothesis which has already been discounted by previous methanol surveys
\citep{phi98,ell94a}, but this method for constraining methanol emission
is a natural way to incorporate the available information about molecular
masing regions in the sample galaxies.

The limits on the 6.7 GHz methanol emission in the survey sample are
listed in Table \ref{targets}:  $t_{on}$ is the on-source integration
time in minutes, rms$_{CH_3OH}$ is the rms noise in a 
6-channel (146 kHz) boxcar averaged spectrum, and $\log L_{CH_3OH}$
is the limit on the methanol luminosity matched to the OH line width.

\section{Discussion}\label{discussion}
\begin{figure}[!p]
\epsscale{0.9}
\plotone{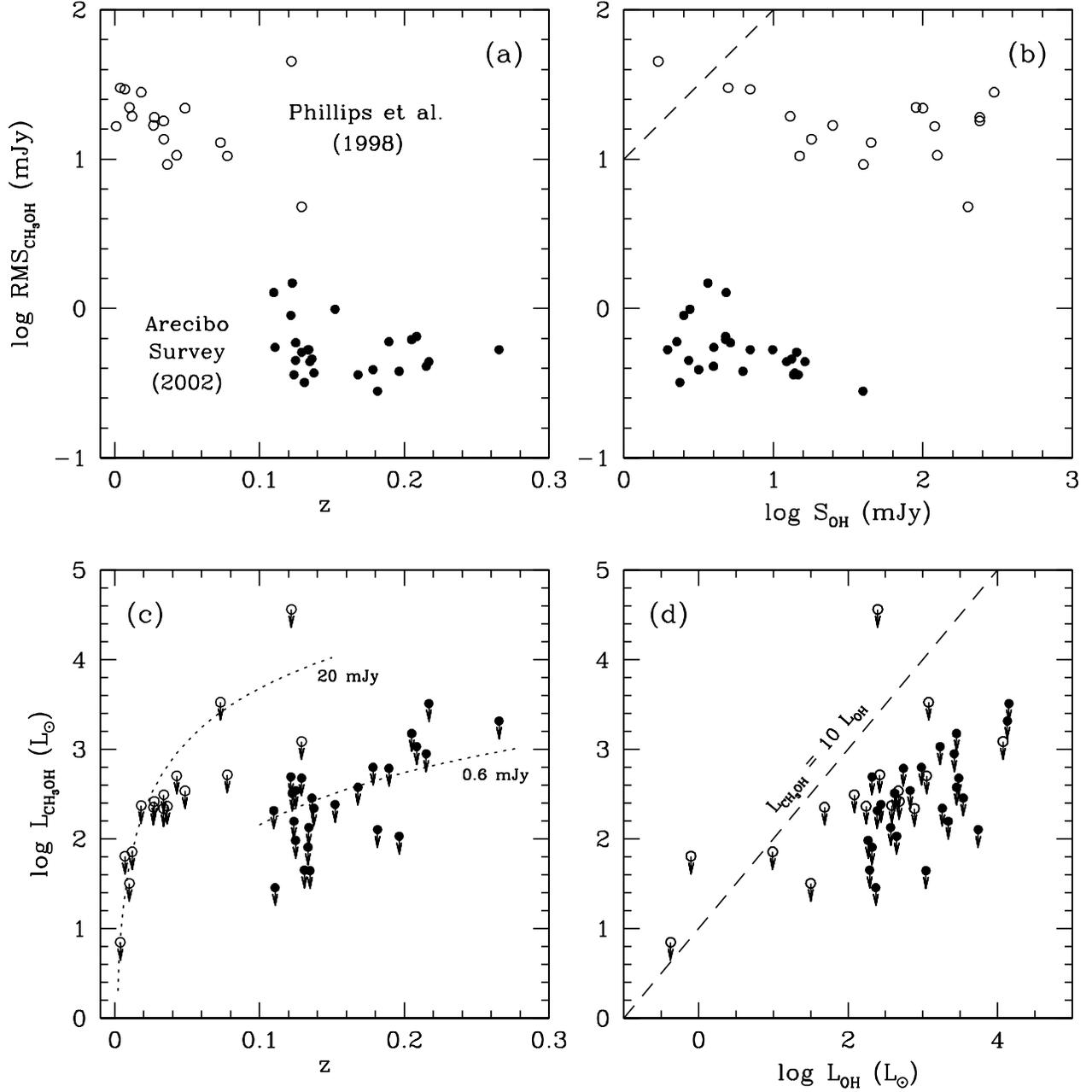}
\caption{Limits on the 6.7 GHz CH$_3$OH emission in OH megamaser galaxies.
Panels
(a) and (b) show the 6.7 GHz rms values obtained by this work and 
by \citet{phi98} versus redshift and OH peak flux density, respectively.  
Filled circles are obtained from this work and open circles are 
from \citet{phi98}.
Panels (c) and (d) plot upper 
limits on the 6.7 GHz isotropic integrated line luminosity (see text)
versus redshift and OH line luminosity.  The dotted lines in panel (c) 
indicate rough rms noise values in 150 kHz channels for the
two surveys assuming rest frame line widths of 150 km s\minusone.  Arrows
indicate upper limits and the long dashed lines indicate the factor of 10
enhancement of methanol emission over OH emission typical in Galactic 
\HII regions.
\label{fig}}
\end{figure}

Figure \ref{fig} illustrates the methanol-OH megamaser parameter space 
excluded by this work and the \citet{phi98} survey.  Although the Arecibo
methanol survey is more sensitive than the \citet{phi98} survey by 
$\sim1.5$ dex, there is significant overlap in the 6.7 GHz line luminosity
limits because the Arecibo survey is on average a factor of 3 higher
in redshift.  
The long-dashed line in panels (b) and (d) of Figure \ref{fig} indicates
the typical order of magnitude methanol to OH intensity ratio found in 
Galactic compact \HII\ regions \citep{cas95a}.  Note 
that this relationship is derived from the 1665 MHz OH line whereas in 
OHMs the 1667 MHz line dominates.  
All of the upper limits on 
the 6.7 GHz methanol emission fall below this relationship.  
The discrepancy is most
significant for the largest $L_{OH}$ objects, indicating a significant 
difference between the physical conditions producing OH megamasers and 
those responsible for Galactic OH masers.  

There are other indications that OHMs represent a fundamentally different 
phenomenon from OH masing associated with compact \HII\  regions.
Foremost of these is the 1667:1665 MHz hyperfine ratio:  in OHMs the
1667 MHz line dominates while in compact \HII\  regions the 1665 MHz
line dominates.  This suggests a lower density environment for OHMs 
\citep{pav96}.
Second, a Galactic star formation rate scaled up to 100--1000 
$M_\odot$ yr\minusone\ 
does not produce the observed OHM line luminosities found in 
(U)LIRGs and cannot explain the OHM fraction in these galaxies or
the dependence of this fraction on the far-IR luminosity \citep{dar02a}.  

In light of the apparent differences between Galactic OH masers and OHMs, 
it may not be surprising that no methanol megamasers have been detected 
in several surveys.  But most of the conditions required for methanol 
masers are still present in OHMs:  adequate molecular columns with coherent 
velocity paths, a strong far-IR radiation bath, and copious continuum 
photons to amplify.  The intensity of 100 and 60 $\mu$m radiation
is more than adequate to pump methanol, even with low efficiency \citep{vdW95}.
What component required for methanol masing is missing in OHM galaxies?  
As suggested by \citet{phi98}, it may be the large-scale presence of 
methanol itself.  Methanol in giant molecular clouds or circumnuclear
molecular tori may be depleted onto dust grains and restricted to 
star formation regions where it is liberated by grain heating.  
In a study of ultracompact \HII\  regions, 
\citet{wal97} find that the number of 6.7 GHz methanol maser spots increases
with peak maser luminosity, which is interpreted to indicate that
maser emission depends on the abundance of methanol rather than on the
far-IR radiation field.  
If methanol megamasers are abundance-limited
rather than pump-limited, then we suggest that
in a dense, massive starburst environment it is possible for the 
output from the occasional ``normal'' methanol maser to be amplified 
by another cloud with inverted methanol level populations.  The required 
velocity coincidence along with geometrical constraints probably 
conspire to make this a rare occurrence, but one that could 
occasionally be found in the nuclei of OHM and luminous IR galaxies.  
Although 
no detections were made in this survey, a survey sensitive to $\sim1$ km 
s\minusone\ lines of a larger sample of IR galaxies may detect a few of
these rare events.  

Numerical modelling can address the absence of 6668 MHz methanol megamasers
in OHM galaxies by extending the detailed combined methanol-OH maser models
of massive star forming regions by Cragg, Sobolev, \& Godfrey (2002) 
to physical conditions 
more appropriate for OHMs.  \citet{kyl98}, \citet{bur90}, and \citet{ran95} 
suggest (with significant disagreement)
$n_{H_2} = 10^3$--$10^5$ cm$^{-3}$, 
$n_{OH}/n_{H_2} = 10^{-6}$--$10^{-8}$,
$T_{dust} \sim 60$ K, 
$T_{kin} = 40$--60 K, 
$L=1$--10 pc, and
$\Delta\mbox{v} = 1$--100 km s\minusone, 
where $L$ is the masing path length and $\Delta\mbox{v}$ is the turbulent
velocity width in the masing path.  The methanol density is a free parameter.
Extrapolation of the \citet{cra02} simulations to likely OHM settings
suggests that methanol would not be detected by our survey if 
$n_{CH_3OH} \leq n_{OH}$ or $T_{dust} < 110$ K.  
While the scaling of $L$, $n_{H_2}$, and 
$\Delta\mbox{v}$ from the values assumed by Cragg \etal
($L = 0.03$ pc, $n_{H_2} > 10^4$ cm$^{-3}$, and $\Delta\mbox{v} = 1$ km 
s\minusone) to OHM conditions 
nearly cancel in the computation of the
OH and methanol specific column density ($N/\Delta\mbox{v}$),
$T_{dust}<110$ K severely quenches methanol masing.  In fact, extrapolation of
the Cragg \etal models to $T_{dust}<100$ K indicates 
that no enhancement of methanol abundance or path length can counteract the
sensitivity of masing methanol lines to the temperature of the IR
radiation field.  Global dust temperatures in (U)LIRGs are generally 30--40 K 
\citep{dun00}, but
they may be significantly higher in small-scale star forming 
regions or in the vicinity of enshrouded AGN.  Although low 
dust temperature
is an attractive explanation for the lack of methanol megamasers, confirmation
would require measurements of $T_{dust}$ in the regions actually producing
OH megamasers rather than a global temperature measurement.  
Note that the 1665 MHz line is always dominant over the 1667 MHz line in 
the \citet{cra02} models whereas the reverse is observed in OHMs.

\section{Conclusions}\label{conclusions}

A sample of OH megamasers at $z\geq0.11$ spanning a wide range of host
galaxy types and OH emission properties has been observed for 
6.7 GHz methanol megamaser emission.  No lines were identified in the
25 RFI-free OHM galaxies within $\pm1750$ km s\minusone\ of the systemic
redshift, and the observations were sensitive to lines from a few to
$\sim2000$ km s\minusone\ in width.  We set limits on methanol emission 
well below the Galactic methanol-OH intensity ratio 
of $\sim10$, although it should be noted that the Galactic ratio uses
the 1665 MHz OH line whereas in OHMs the 1667 MHz line dominates.  Combined
with 17 OHMs observed by \citet{phi98}, this study supports the notion
that OH megamasing is a fundamentally different process from the OH masing
associated with Galactic compact \HII\  regions and methanol masers.  
Maser models suggest that 6668 MHz methanol masers are strongly quenched 
in OH megamaser galaxies in which $T_{dust}<100$ K or 
$n_{CH_3OH} \leq n_{OH}$.

\acknowledgements

The authors wish to thank the excellent staff of NAIC for observing assistance
and support, especially Karen O'Neil, and the AIPS++ consortium for software
support, especially Joe McMullin.  
We thank the anonymous referee for an excellent suggestion and thoughtful
comments.
This research was supported by NSF grant AST 00-98526 and made
use of the NASA/IPAC Extragalactic Database (NED) 
which is operated by the Jet Propulsion Laboratory, California
Institute of Technology, under contract with the National Aeronautics 
and Space Administration.  The National Astronomy and Ionosphere Center 
is operated by Cornell University
under a cooperative agreement with the National Science Foundation.

\begin{deluxetable}{cccccccccc}
\tablecaption{OH Megamasers Surveyed for Methanol 6.7 GHz Emission\label{targets}}
\tablewidth{0pt}
\tablehead{
\colhead{{\it IRAS} Name} &  
\colhead{$\alpha$} & 
\colhead{$\delta$} & 
\colhead{$z$} & 
\colhead{$S_{OH}$} &
\colhead{$\Delta$v$_{OH}$} & 
\colhead{$\log L_{OH}$} &
\colhead{$t_{on}$} &
\colhead{rms$_{CH_3OH}$} &
\colhead{$\log L_{CH_3OH}$} \\
\colhead{FSC} &\colhead{B1950} & \colhead{B1950} & \colhead{} & \colhead{(mJy)}
& \colhead{(km s\minusone)} & \colhead{($h^{-2}_{75} L_\odot$)} 
& \colhead{(min)} & \colhead{(mJy)}
& \colhead{($h^{-2}_{75} L_\odot$)} \\
\colhead{(1)}&\colhead{(2)}&\colhead{(3)}&\colhead{(4)}&\colhead{(5)}&
\colhead{(6)}&\colhead{(7)}&\colhead{(8)}&\colhead{(9)}
}
\startdata
02524+2046 & 02 52 26.8 & +20 46 54 & 0.1815 & 39.82 & \phn76 & 3.71 & 35 & 0.28 & $<2.1$ \\
03521+0028 & 03 52 08.5 & +00 28 21 & 0.1522 & \phn2.77 & \phn59 & 2.41 & 25 & 0.99 & $<2.4$ \\
03566+1647 & 03 56 37.8 & +16 47 57 & 0.1335 & \phn1.96 & \phn48 & 2.29 & 25 & 0.53 & $<1.9$ \\
04121+0223 & 04 12 10.5 & +02 23 12 & 0.1216 & \phn2.52 & 209 & 2.30 & 20 & 0.90 & $<2.7$ \\
06487+2208 & 06 48 45.1 & +22 08 06 & 0.1437 & \phn7.60 &    211 & 2.86 & \nodata & RFI & \nodata\\
07163+0817 & 07 16 23.7 & +08 17 34 & 0.1107 & 4.00 & \phn24 & 2.35 & 35 & 0.55 & $<1.4$ \\
07572+0533 & 07 57 17.9 & +05 33 16 & 0.1894 & \phn2.26 & 156 & 2.71 & 25 & 0.60 & $<2.8$ \\
08201+2801 & 08 20 10.1 & +28 01 19 & 0.1680 & 14.67 & 205 & 3.42 & 25 & 0.36 & $<2.5$ \\
08279+0956 & 08 27 56.1 & +09 56 41 & 0.2085 & \phn4.79 & 207 & 3.19 & 25 & 0.65 & $<3.0$ \\
09039+0503 & 09 03 56.4 & +05 03 28 & 0.1250 &  \phn5.17 & 212 & 2.80 & 25 & 0.59 & $<2.5$ \\
09531+1430 & 09 53 08.3 & +14 30 22 & 0.2151 &  \phn3.98 & 256 & 3.38 & 25 & 0.41 & $<2.9$ \\
09539+0857 & 09 53 54.9 & +08 57 23 & 0.1290 &  14.32 & 317 & 3.45 & 25 & 0.51 & $<2.7$ \\
10339+1548 & 10 33 58.1 & +15 48 11 & 0.1965 &  \phn6.26 &\phn40& 2.62 & 25 & 0.38 & $<2.0$ \\
10378+1108 & 10 37 49.1 & +11 09 08 & 0.1362 & 	13.3\phn & 188 & 3.26 & 25 & 0.46 & $<2.4$ \\
11524+1058 & 11 52 29.6 & +10 58 22 & 0.1784 & \phn3.17 & 279 & 2.95 & 25 & 0.39 & $<2.8$ \\
12005+0009 & 12 00 30.2 & +00 09 24 & 0.1226 &  \phn3.65 &\phn82 & 2.61 & 15 & 1.48 & $<2.5$ \\
12032+1707 & 12 03 14.9 & +17 07 48 & 0.2170 & 16.27 & 853 & 4.11 & 20 & 0.44 & $<3.5$ \\
12162+1047 & 12 16 13.9 & +10 47 58 & 0.1465 &  \phn2.19 & 105 & 2.25 & \nodata & RFI & \nodata\\
13218+0552 & 13 21 48.4 & +05 52 40 & 0.2051 &  \phn4.81 & 314 & 3.49 & 25 & 0.62 & $<3.1$ \\
14059+2000 & 14 05 56.4 & +20 00 42 & 0.1237 &  13.65 & 161 & 3.28 & 25 & 0.36 & $<2.2$ \\
14070+0525 & 14 07 00.3 & +05 25 40 & 0.2655 &  \phn9.9\phn & 298 & 4.11 & 25 & 0.53 & $<3.3$ \\
14553+1245 & 14 55 19.1 & +12 45 21 & 0.1249 &  \phn2.72 &\phn77 & 2.21 & 25 & 0.45 & $<2.0$ \\
14586+1432 & 14 58 41.6 & +14 31 53 & 0.1477 & \phn7.11 & 369 & 3.38 & \nodata & RFI & \nodata\\
15224+1033 & 15 22 27.4 & +10 33 17 & 0.1348 & 12.27 &\phn31 & 3.01 & 25 & 0.44 & $<1.6$ \\
15587+1609 & 15 58 45.5 & +16 09 23 & 0.1375 & 13.91 & 176 & 3.23 & 25 & 0.37 & $<2.3$ \\
16100+2527 & 16 10 00.4 & +25 28 02 & 0.1310 &  \phn2.37 &\phn46 & 2.26 & 30 & 0.32 & $<1.6$ \\
16255+2801 & 16 25 34.0 & +28 01 32 & 0.1340 &  \phn7.02 &\phn79 & 2.54 & 20 & 0.53 & $<2.1$ \\
17161+2006 & 17 16 05.8 & +20 06 04 & 0.1098 &  \phn4.84 & \phn76 & 2.37 & 5 & 1.28 & $<2.3$ \\
\enddata 
\end{deluxetable}

\end{document}